# Balancing Higher Moments Matters for Causal Estimation: Further Context for the Results of Setodji et al. (2017)

Melody Y. Huang — University of California, Los Angeles
Brian G. Vegetabile — RAND Corporation
Lane F. Burgette — RAND Corporation
Claude Setodji — RAND Corporation
Beth Ann Griffin — RAND Corporation

July 2021

## Abstract

We expand upon the simulation study of Setodji et al. (2017) which compared three promising balancing methods when assessing the average treatment effect on the treated for binary treatments: generalized boosted models (GBM), covariate-balancing propensity scores (CBPS), and entropy balance (EB). The study showed that GBM can outperform CBPS and EB when there are likely to be non-linear associations in both the treatment assignment and outcome models and CBPS and EB are fine-tuned to obtain balance only on first order moments. We explore the potential benefit of using higher-order moments in the balancing conditions for CBPS and EB. Our findings showcase that CBPS and EB should, by default, include higher order moments and that focusing only on first moments can result in substantial bias in both CBPS and EB estimated treatment effect estimates that could be avoided by the use of higher moments.

# Introduction

A recent paper presented extensive simulation studies aiming to help applied researchers identify when to choose between two competing approaches of methods that create weights to perform weighted causal estimation: 1) methods that directly optimize covariate balance explicitly as constraints in an optimization procedure such as the covariate balancing PS (CBPS)[1] or entropy balancing (EB)[2] and 2) generalized boosted model PSs (GBM).[3] In theory, direct covariate balance results in exact balance between the exposed and control arms for specified moments of potential confounding variables. Alternatively, estimating PSs via GBM[3] uses a flexible, tree-based model to estimate PSs and only uses covariate balance to select a tuning parameter that determines the model complexity and therefore will not generally result in exact balance between treated and control arms. A key finding of simulations in Setodji et al. (2017) was that CBPS and EB perform very well when linearity assumptions hold for either the PS or outcome model, but that GBM performed better in settings where there were substantial non-linearities in both the PS and outcome models.[4] Similarly, a recent simulation study by Li and Li (2021) demonstrated that EB and CBPS performed poorly in non-linear settings, warning researchers that exact balancing approaches should only be used when the PS or outcome model is known with high confidence.[5] A limitation of these past studies is that the direct balancing algorithms only optimized first-moments, i.e., only constraining the mean when optimizing weights for balance, which is the default behavior of most popular software packages for these approaches.

In this brief report, we expand the simulation study of Setodji et al. (2017) to include higher-order moments in the balancing conditions for CBPS and EB. Rather than merely requiring that the mean of each covariate be the same in the treatment and control arms, we additionally require balance on higher moments (i.e., ensure equal variance and higher moments by including square and cubic power functions of covariates in the balancing conditions). Although this is not the default behavior of most software implementations, our simulations show that focusing only on first moments can result in substantial bias in both CBPS and EB in the treatment effect estimates that could be avoided by the use of higher moments. Moreover, using higher moments in EB results in weights that consistently perform better than GBM under the data generation of Setodji et al. (2017).

**Simulation**

We utilize the same simulation design as Setodji et al. (2017).[4] The simulation seeks to test the performance of the different estimators under varying levels of non-linearities in the outcome model and the treatment assignment mechanism. For full details, please refer to the original Setodji et al. (2017) description.[4] To summarize, the simulation tested seven different treatment assignment mechanisms and five outcome models, all with varying degrees of non-linear and non-additive terms. Additionally, six different estimation strategies were tested, in which each strategy utilized a different set of variables to estimate the PS models. In total, there are 35 different scenarios, and six different estimation strategies. For each version of the simulation, 1,000 iterations were run.

To estimate the PS models, we fit four different models: a standard logistic regression, CBPS, EB, and GBM. For CBPS, we used two different methods to estimate the weights. The first uses an over-identified model combining both the PS and covariate balancing conditions, while the second identifies weights using only the covariate balancing conditions. Additionally, for CBPS and EB, we generated three sets of weights: one estimated using just the first moments, one using the first and second moments, and finally, one estimated using all three moments. Finally, for GBM, the weights were first estimated using the default settings from the TWANG package. After obtaining the weights, a stabilized-PS weighted estimator was used to estimate the average treatment effect across the simulation.[5]

## Results

### *Bias Reduction from Including Higher Order Moments*

In the original simulation study, Setodji et al. (2017) highlighted that CBPS did not perform as well as GBM in non-linear scenarios.[4] For example, under moderate amounts of non-linear and non-additive terms, the CBPS estimated weighted estimator had roughly 67% more bias on average than an estimator using GBM-estimated weights. Similarly, using EB failed to provide much more additional bias reduction over CBPS. The highlighted advantage to using a boosted model, such as GBM, comes from the fact that the tree-based models are able to capture complex non-linear and interactive relationships.

However, when we add in higher moment balancing conditions to CBPS and EB, there is a large degree of bias reduction from simply including the second order moment (see Table 1). In the case of CBPS, the absolute bias across simulations for a scenario with non-linear and non-additive terms is reduced from 0.48 to around 0.16. For EB, the absolute bias is reduced from 0.36 to 0.09 . Thus, the inclusion of the second order moments results in a 50-60% reduction in absolute bias for both CBPS and EB. When adding in third order moments, the absolute bias from EB weights is reduced further to 0.06, but the majority of the bias reduction appears to be coming from the inclusion of the second order moment. The absolute bias GBM is in this scenario is 0.38; thus, by including the second or third order moments, the absolute bias for both CBPS and EB will be greatly reduced compared to GBM.

*Robustness of Balancing on Higher-Order Terms*

A drawback to GBM that the original Setodji et al. (2017) study showcases is that in cases when the underlying PS model is truly linear (i.e., Scenario A), GBM incurs more relative bias than a standard logistic regression.[4] This is due to the fact that the tree-based structure imposes piece-wise constants to estimate the PS within regions of the covariate space that can be difficult to model linear functions (i.e., performing best when non-linearities are present). Therefore, practitioners would have to determine whether they believe the underlying treatment assignment mechanism is inherently linear or non-linear before proceeding with their weight estimation. However, we see that for both CBPS and EB, in linear scenarios, while balancing on the first moment is sufficient to reduce bias, adding in the higher order moments to approximate potential non-linearities, even when none exist, does not lead to worse bias. Thus, balancing on higher moments may offer protection in the case that the problem is non-linear in nature, but does little to harm inferences when linear assumptions suffice.

*Additional Notes*

In the original Setodji et al. (2017) study, CBPS was estimated using the default parameters.[4] By default, CBPS will fit an over-identified model, using both PS and covariate balancing conditions.[1] In general, we observe that EB provides slightly more bias reduction over this default specification of CBPS. An alternative approach is to estimate CBPS using only covariate balancing. We also include this version of CBPS and, as expected, it performs more similarly to EB. However, unlike EB, one advantage of CBPS is that it provides direct estimates for PSs. In practice, it can be useful to have access to the estimated PSs for conducting sensitivity analysis. We highlight that using the modified version of CBPS can allow for an alternative way to estimate weights that deliver similar performance to EB, with the added advantage of having estimated PSs.

## Conclusion

Previous studies found that GBM can substantially outperform CBPS or EB when there are likely to be non-linear associations in both the PS and outcome models. However, including higher order moments in CBPS and EB-type weighting approaches can provide more robust estimates even when there are substantial non-linear or interactive effects. These simulations suggest that CBPS and EB should, by default, include higher order moments, and that the range of situations for which GBM can outperform properly implemented CBPS or EB is narrower than previously believed. The findings regarding the benefits of using higher order moments are critically needed for the field, given the popularity of both CBPS and EB and the fact that the default behavior of most software implementations is to solely match on the first moments for both methods. Our simulations clearly show that focusing only on first moments can result in substantial bias in both CBPS and EB estimated treatment effect estimates that could be avoided by the use of higher moments.

# Tables

**Table 1: Summary of estimator performances across the different simulations.**

| | Absolute Bias | | | | | | | | | | | RMSE | | | | | | | | | | |
|---|---|---|---|---|---|---|---|---|---|---|---|---|---|---|---|---|---|---|---|---|---|---|
| | | | Entropy Balancing | | | CBPS (Default) | | | CBPS (Exact) | | | | | Entropy Balancing | | | CBPS (Default) | | | CBPS (Exact) | | |
| PS | Logit | GBM | $m=1$ | $m=2$ | $m=3$ | $m=1$ | $m=2$ | $m=3$ | $m=1$ | $m=2$ | $m=3$ | Logit | GBM | $m=1$ | $m=2$ | $m=3$ | $m=1$ | $m=2$ | $m=3$ | $m=1$ | $m=2$ | $m=3$ |
| A | 0.13 | 0.24 | 0.11 | 0.08 | 0.06 | 0.15 | 0.14 | 0.14 | 0.12 | 0.08 | 0.06 | 0.07 | 0.11 | 0.06 | 0.05 | 0.03 | 0.08 | 0.07 | 0.07 | 0.07 | 0.05 | 0.03 |
| B | 0.14 | 0.23 | 0.12 | 0.08 | 0.06 | 0.13 | 0.13 | 0.13 | 0.13 | 0.09 | 0.06 | 0.07 | 0.10 | 0.06 | 0.04 | 0.03 | 0.07 | 0.07 | 0.06 | 0.07 | 0.05 | 0.03 |
| C | 0.38 | 0.38 | 0.41 | 0.08 | 0.07 | 0.49 | 0.20 | 0.20 | 0.40 | 0.10 | 0.07 | 0.16 | 0.16 | 0.17 | 0.04 | 0.03 | 0.20 | 0.09 | 0.09 | 0.17 | 0.05 | 0.04 |
| D | 0.14 | 0.27 | 0.12 | 0.09 | 0.06 | 0.15 | 0.15 | 0.15 | 0.13 | 0.08 | 0.06 | 0.08 | 0.12 | 0.06 | 0.05 | 0.03 | 0.08 | 0.07 | 0.07 | 0.07 | 0.05 | 0.03 |
| E | 0.16 | 0.26 | 0.14 | 0.09 | 0.06 | 0.14 | 0.15 | 0.15 | 0.14 | 0.09 | 0.06 | 0.09 | 0.11 | 0.07 | 0.05 | 0.03 | 0.07 | 0.07 | 0.07 | 0.08 | 0.05 | 0.03 |
| F | 0.15 | 0.25 | 0.13 | 0.10 | 0.07 | 0.13 | 0.12 | 0.13 | 0.13 | 0.09 | 0.07 | 0.08 | 0.11 | 0.07 | 0.05 | 0.04 | 0.07 | 0.06 | 0.06 | 0.07 | 0.05 | 0.04 |
| G | 0.27 | 0.38 | 0.36 | 0.09 | 0.06 | 0.48 | 0.16 | 0.18 | 0.36 | 0.09 | 0.07 | 0.12 | 0.16 | 0.16 | 0.05 | 0.03 | 0.20 | 0.08 | 0.08 | 0.15 | 0.05 | 0.04 |

Footnotes for Table:
We see a significant reduction in bias in the balancing approaches from including higher order moments. Furthermore, the overall RMSE is also lower from including higher order moments.

Treatment Assignment Mechanisms:
  A. Additive and linear:
  $$\text{Version}_A = \alpha_1 X_1 + \alpha_2 X_2 + \alpha_3 X_3 + \alpha_4 X_4 + \alpha_5 X_5 + \alpha_6 X_6 + \alpha_7 X_7$$
  B. Mild non-linearity:
  $$\text{Version}_B = \text{Version}_A + \alpha_8 X_2^2$$
  C. Moderate non-linearity:
  $$\text{Version}_C = \text{Version}_A + \alpha_9 X_2^2 + \alpha_{10} X_4^2 + \alpha_{11} X_7^2$$
  D. Mild non-additivity:
  $$\text{Version}_D = \text{Version}_A + \alpha_{12} X_1 X_3 + \alpha_{13} X_2 X_4 + \alpha_{14} X_4 X_5 + \alpha_{15} X_5 X_6$$
  E. Mild non-additivity and non-linearity:
  $$\text{Version}_E = \text{Version}_D + \alpha_{16} X_2^2$$
  F. Moderate non-additivity:
  $$\text{Version}_F = \text{Version}_D + \alpha_{17} X_5 X_7 + \alpha_{18} X_1 X_6 + \alpha_{19} X_2 X_3 + \alpha_{20} X_3 X_4 + \alpha_{21} X_4 X_5$$
  G. Moderate non-linearity and non-additivity:
  $$\text{Version}_G = \text{Version}_F + \alpha_{22} X_2^2 + \alpha_{23} X_4^2 + \alpha_{24} X_7^2$$

Outcome Models:
  1. Additive and linear (Main effects terms only):
  $$Y = \delta_0 + \theta T + \delta_1 X_1 + \delta_2 X_2 + \delta_3 X_3 + \delta_4 X_4 + \delta_5 X_8 + \delta_7 X_{10}$$
  2. Mild non-linearity:
  $$Y = \delta_0 + \theta T + \delta_1 X_1 + \delta_2 X_2^2 + \delta_3 X_3 + \delta_4 \exp(1.3 \cdot X_4) + \delta_5 X_8 + \delta_6 X_9 + \delta_7 X_{10}$$
  3. Moderate non-linearity
  $$Y = \delta_0 + \theta T + \exp(\delta_1 X_1 + \delta_2 X_2 + \delta_3 X_3 + \delta_4 X_4 + \delta_5 X_8 + \delta_6 X_9 + \delta_7 X_{10})$$
  4. Severe non-linearity:
  $$Y = \delta_0 + \theta T + \exp(\delta_1 X_1 + \delta_2 X_2 + \delta_3 X_3) + \delta_4 \exp(1.3 \cdot X_4) + \delta_5 X_8 + \delta_6 X_9 + \delta_7 X_{10}$$
  5. Sinusoidal non-linearity:
  $$Y = \delta_0 + \theta T + 4\sin(\delta_1 X_1 + \delta_2 X_2 + \delta_3 X_3 + \delta_4 X_4 + \delta_5 X_8 + \delta_6 X_9 + \delta_7 X_{10})$$

Estimation Strategy:
  1. Only confounders directly related to both the outcome and treatment
  2. Variables only directly associated with the treatment, including the instrument $X_7$
  3. Only variables associated directly to the outcome
  4. All variables directly or indirectly related to both outcome and treatment
  5. All variables related to the outcome or to the treatment
  6. All available variables including distractors

# References

1. Imai K, Ratkovic M. Covariate balancing propensity score. *Journal of the Royal Statistical Society: Series B: Statistical Methodology.* 2014: 243–263.

2. Hainmueller J. Entropy balancing for causal effects: A multivariate reweighting method to produce balanced samples in observational studies. *Political Analysis.* 2012: 25–46.

3. McCaffrey DF, Ridgeway G, Morral AR. Propensity score estimation with boosted regression for evaluating causal effects in observational studies. *Psychological Methods.* 2004;9:403.

4. Setodji CM, McCaffrey DF, Burgette LF, Almirall D, Griffin BA. The right tool for the job: Choosing between covariate balancing and generalized boosted model propensity scores. *Epidemiology.* 2017;28:802.

5. Li Y, Li L. Propensity score analysis methods with balancing constraints: A Monte Carlo study. *Statistical Methods in Medical Research*. 2021;30:1119-1142.

6. Särndal CE, Swensson B, Wretman J. *Model Assisted Survey Sampling.* New York: Springer-Verlag; 2003.